# Electrochemical Glucose Sensor using Single-Wall Carbon Nanotube Field Effect Transistor

Reetu Raj Pandey[1*], Jie Liang[1], Dilek Cakiroglu[2], Benoit Charlot[2], Aida Todri-Sanial[1*]

1   LIRMM, UMR CNRS 5506, University of Montpellier, 34095 Montpellier Cedex 5, France
2   IES, UMR CNRS 5214, University of Montpellier, 34095 Montpellier Cedex 5, France
*Corresponding email: aida.todri@lirmm.fr, pandey@lirmm.fr

*Abstract*—In this paper, we present a simple yet sensitive method for glucose sensing using carbon nanotube field-effect transistor (CNTFET) based biosensor. The CNTs were well-dispersed to form CNT networks and maintain functional connectivity among CNTs, which increases the electron transfer through the network and thus, the electronic readout. Moreover, glucose oxidase (GOx) molecules are immobilized by CNT functionalization to form effective and sensitive CNT networks as FET channel. The CNTs are functionalized with linkers (1-pyrenebutanoic acid succinimidyl ester) to immobilize GOx on CNTs, where GOx serves as a mediator between CNTs and glucose for electron transfer. The liquid analyte glucose is adsorbed on CNTs via GOx and linkers by releasing additional electrons in the CNTFET channel and thus, increasing the CNTFET readout current. The binding of the target glucose molecules and GOx emulates the gate potential of FET channel and the electronic response of the sensor is recorded in real-time. Moreover, the variations in electronic readout of CNTFET biosensor are observed and is stipulated due to variation in CNT dispersion on each device. Overall, this work presents a simple, fast, sensitive, low-cost, and low concentration (0.01 mM) detection of glucose using CNTFET sensors.

**Introduction**

Carbon nanotubes (CNTs) has gained considerable interest in biosensing applications due to its unique electrical, physical and optical properties [1-3]. More importantly, the well-known carbon chemistry, easy fabrication steps and the high surface sensitivity makes it more significant for biosensors [4-6]. The functionalized CNTs with its intrinsic high electron-transfer rate and fast electrocatalytic effects are highly desirable for biomolecular detections [7, 8]. Moreover, field-effect transistor (FET) based sensors have gained much interest by providing a means of estimating the analyte concentration by measuring the FET channel current of enzymatically liberated charges on the CNT surface [9-13]. Thus, several groups have explored CNT-FET sensors where the solid-state gate is replaced by organic molecules adsorbed on the channel, which modulates the channel conductance [14-16]. Furthermore, some reports had proposed single-wall carbon nanotubes (SWCNTs) with its semiconducting properties is a promising candidate for biosensors [17, 18]. The first SWCNT based chemical sensor was proposed by Kong et al. in 2000 for gas detection [16] and Chen et al. [19] were the first to immobilize proteins on SWCNTs. These proteins and enzymes work as an electrostatic gate on SWCNT channel and the gating effect modulates the channel conductance improving the sensitivity. Several works on enzymatic and non-enzymatic sensors were already reported [14, 20, 21] and some of them suggest that the non-enzymatic sensors have larger sensitivity, as there is a direct electron transfer from the analyte to CNT [22, 23]. However, the complexity of non-enzymatic sensors is the principal drawback for good sensor design, since they use additional nanoparticles on the CNT to react with the analyte and increase the sensitivity. Thus, despite the tremendous progress in enzymatic and non-enzymatic glucose sensor, we propose an alternative, simple, fast yet sensitive CNT-FET based glucose sensor with a detection limit of 0.01 mM glucose.

Here, we present SWCNTFET (namely CNTFET hereafter) electrochemical biosensor for glucose detection. The CNTFET sensors are investigated where the FET gate voltage is emulated by the charges (e.g. organic molecules) adsorbed on the FET channel to promote an efficient electron transfer. It can be seen from Fig. 1(a), CNTs are homogeneously dispersed on Si/SiO2 substrate with a set of interdigitated electrodes on it. The CNT network with its high surface to volume ratio enables to immobilize a large number of linking molecules on CNTs, thus, increasing the electron transfer between the analyte and CNTs and increase in current (i.e., sensor sensitivity). The immobilization of proteins on SWCNTs is shown where these proteins acted as an electrostatic gate on SWCNT channel and enabled tuning of channel conductance. The change in source-drain current is measured as a function of analyte concentration to obtain the sensitivity of the sensor.

**Experimental Method**

CNT-FET devices were fabricated using the conventional lithography process and then functionalized to immobilize the biomolecules on the CNTs. We used commercial SWCNTs from Sigma Aldrich with ~98% semiconducting purity. The average diameter of SWCNTs is 1.2 nm and length between 0.5 to 5 µm. The solution of SWCNTs in ethanol was prepared with 0.05 mg/50 ml concentration and are sonicated with a bar type sonicator for 4 hr. Dispersed SWCNTs were left for 10 mins and then drop-cast on Si/SiO2 wafer at 40℃, which is the optimal temperature observed to obtain a well-dispersed network of SWCNTs. At this temperature, we found that ethanol evaporates without forming CNT agglomerations or capillary aggregations such as the coffee rings effect.

Metal electrodes were fabricated on top of dispersed CNTs network with a conventional photolithography process. Further, thin layers of 5 nm chromium followed by 50 nm gold were deposited on the patterned electrodes, and lift-off is followed to remove the extra metals for final electrodes. The distance between the interdigitated electrodes is 8 µm with a length of 4.5 mm and a width of 2 µm. The schematic of the fabricated device is shown in Fig. 1(a) and the SEM images are shown in Fig. 2.

Once the CNTFET devices were ready, the solvent of 1-pyrenebutanoic acid succinimidyl ester in dimethylformamide (DMF) was prepared by stirring and mixing with a concentration of 2.3 mg/ml for CNT functionalization. CNTFET was dipped in the prepared solvent for 2 hrs and then rinsed with clean DMF to remove the extra 1-pyrenebutanoic acid succinimidyl ester adsorbed on the substrate. Immobilization of GOx on the CNTs was obtained by dipping the CNT-FET in GOx solvent for 18 hrs. The concentration of GOx solution was 10mg/ml in de-ionized water. Finally, the sample device was washed within de-ionized water to remove extra GOx adsorbed on the substrate. After each step, the CNTFET devices were characterized by a four-probe system and semiconductor parameter analyzer (Keysight B2901A).

**Results and Discussion**

The pristine CNTFET devices without any functionalization were electrically characterized and electronic responses were recorded. The current-voltage curves of FET show a p-type behaviour as expected and, is due to exposure of oxygen and humidity on the device that act as acceptors [24]. Moreover, devices were tested at each stage of functionalization to determine the change in current-voltage response as shown in Fig. 3(a). In total, 18 devices were tested at each step during the functionalization. The functionalization steps include 1) pristine, 2) immersed in DMF for 2 hrs, 3) dipped in DMF solvent with 2.3 mg/mL 1-pyrenebutanoic acid succinimidyl ester for 2 hrs, and 4) after the GOx immobilization. All 18 devices showed similar behaviour during the functionalization and GOx immobilization. As shown in Fig. 3(a) the drain current was highly suppressed after DMF, which is due to electron donor nature of DMF upon exposure with CNT network [7]. Moreover, the linking molecule (pyrenebutanoic acid succinimidyl ester) had almost no effect on the current-voltage curve. Whereas, GOx immobilization had an impact and the drain current decreased as shown in Fig. 3(a). The decrease in current indicates that the device can measure the presence of GOx and enzyme detection.

In this work, we apply the liquid top gate on functionalized CNTFET devices shown in the schematic Fig. 1(b), as the top gate provides better electrostatic control to the FET channel compared to the back gate [25]. The liquid gate is an efficient gate for glucose sensing as the CNTs are immersed in analyte and sensing is performed in a liquid environment [7]. Ag/AgCl standard electrode was used to apply the liquid top gate voltage on the device. First, the current-voltage (Id-Vds) curves of the devices were measured in the presence of deionized water (without glucose) at fixed Vgs =-1.5 V and then with glucose, as shown in Fig. 3 (b, c and d). Chemical reactions take place where glucose in the presence of $O_2$ reacts with GOx and forms Gluconic acid and $H_2O_2$ as a by-product. Whereas, GOx converts the glucose into gluconolactone with the reduction of GOx.

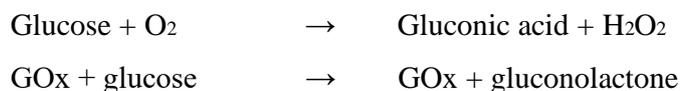

$$\text{Glucose} + O_2 \rightarrow \text{Gluconic acid} + H_2O_2$$
$$\text{GOx} + \text{glucose} \rightarrow \text{GOx} + \text{gluconolactone}$$

We investigate CNTFET device response with various glucose concentrations to obtain the biosensor sensitivity with the lowest glucose concentrations. Sensitivity to low glucose concentrations can be relevant to detect the glucose level, not only in blood but also in other body fluids such as saliva, urine, sweat and tears [26]. Three sets of devices with six devices in each set were prepared to detect glucose concentration of 0.01 mM, 0.05 mM and 0.1 mM, respectively. The change is CNTFET device response with the glucose concentrations is shown in Fig. 3 (b, c and d). The curves in Fig. 3 (b, c and d) are the best case of current variation (i.e., ΔI) out of 3 sets of devices. We quantified electronic response of FET snsor and the current change with and without glucose and listed in Table 1. The sensitivity of these 18 devices with respect to current change and glucose concentration were determinedand we observe that all the six devices in set 1 have different current with the same glucose concentration (0.01 mM). The difference in current is due to CNT network variation from one CNTFET device to another, as the CNT dispersion varies from device to device. The current-voltage curves with Vgs=-1.5 V for all devices can be found in supporting info (S1, S2, S3).

Additionally, a given device with varying concentrations of glucose was also tested, as shown in Fig. 4. We observe that the device reaches saturation after a specific glucose concentration. The device conductance varied, but the sensor showed a linear response from 0.01 to 2 mM and started to saturate after 2 mM. The sensitivity of the sensor was obtained from the slope of the curve in the linear region. The sensitivity of the sensor was in the range of 0.4 to 0.6 µA/mM. The real-time response of the sensor was obtained and shown in Fig. 5(a). There was a sharp increase in current with glucose, which confirms the device sensitivity. The drain-source currents with the various concentration of glucose are plotted in Fig. 5(b). Overall, the CNTFET device is suitable to detect a range of 0.01 to 2 mM glucose concentrations. Such results are encouraging, given that the CNTFET device uses a non-aligned CNT network with large electrodes. Such devices could potentially eliminate the need for aligned CNTs in CNTFET biosensors, which are more complex and require costly fabrication process.

**Conclusion**

CNTs have been demonstrated as a good candidate material to immobilize the GOx enzyme. In this work, we present a low-cost and simple, yet sensitive biosensor based on non-aligned CNT network as a channel material for CNTFET device. CNT networks were fabricated with a simple and straightforward process via sonication and drop cast on the wafer at 40℃. Immobilization of GOx was confirmed with the change in the electrical response of the CNTFET device. The difference in the electrical signal can be observed due to adsorbed molecules on the CNT surface, which changes the conductivity in the FET channel. The obtained CNTFET devices can detect a range of glucose concentrations (0.01 to 2 mM) or sensitivity of 0.4 to 0.6 µA/mM. Such CNTFET devices with non-aligned CNT network show promising properties for measuring enzymatic activity and can be integrated into small chips.

**ACKNOWLEDGMENT**

Authors would like to acknowledge H2020 SmartVista European project. This project has received funding from the European Union's Horizon 2020 research and innovation programme under grant agreement No. 825114 (http://www.smartvista.eu).

**Figures**

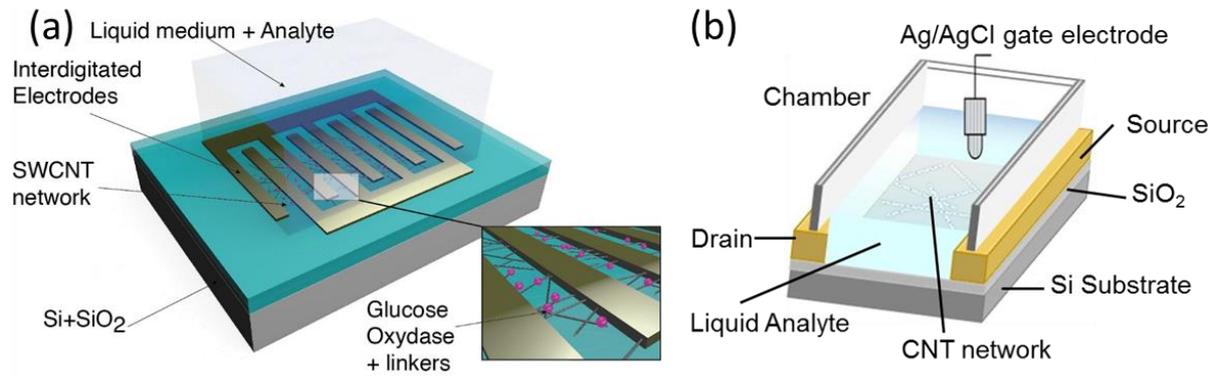

Fig.1. Schematic of fabricated device. (a) Interdigitated electrodes of CNTFET device and its isolation for the liquid medium measurement, (b) Top gate measurement setup illustration.

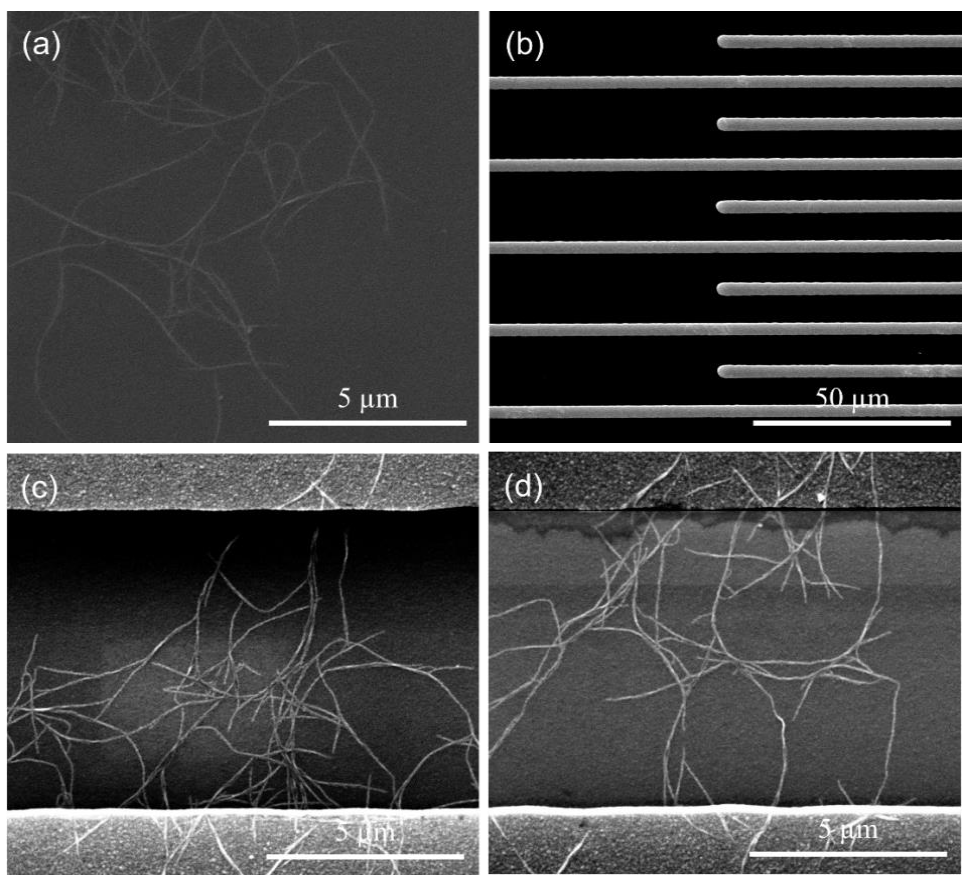

Fig. 2. SEM images of CNT network on substrate. (a) Dispersion of CNTs (b) Interdigitated electrodes to bridge the CNTs. (c, d) CNT network bridged with the electrodes.

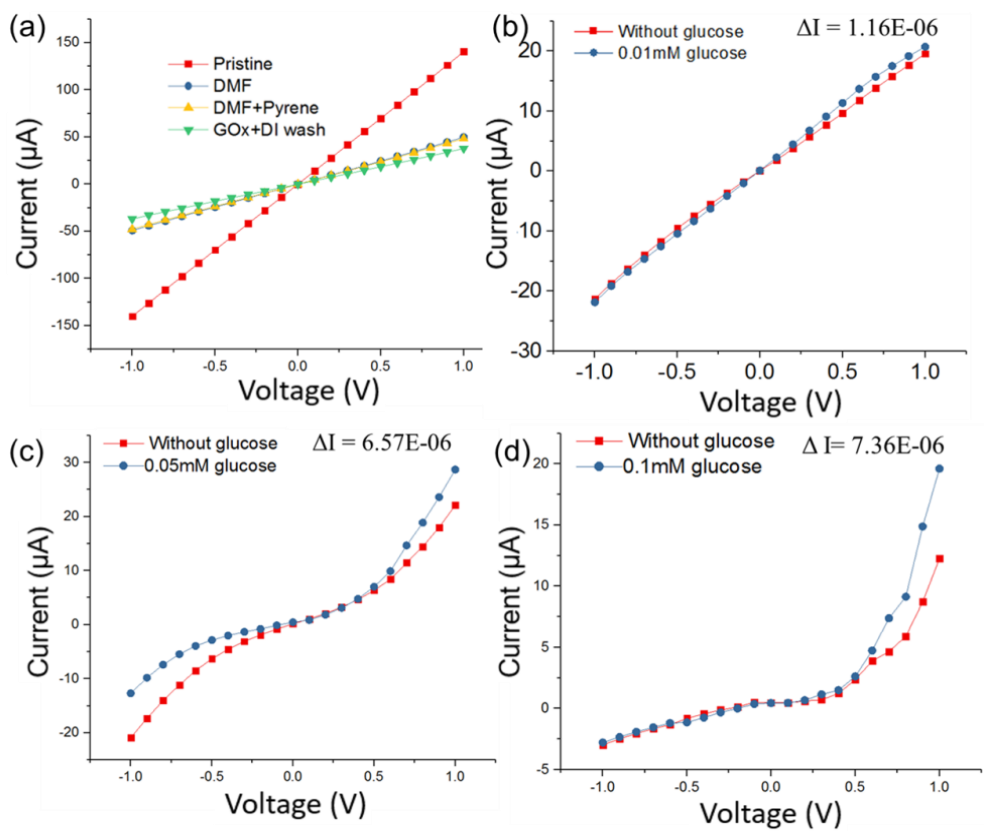

Fig. 3. Electrical characteristics of CNT-FET. (a) Current vs Voltage ($I_{ds}$-$V_{ds}$) of pristine and functionalized device without gate effect. (b, c and d) $I_{ds}$-$V_{ds}$ of electrochemical gated glucose sensing with $V_{gs}$= -1.5 V and varying glucose concentration. The delta change in current is obtained at $V_{ds}$=1V.

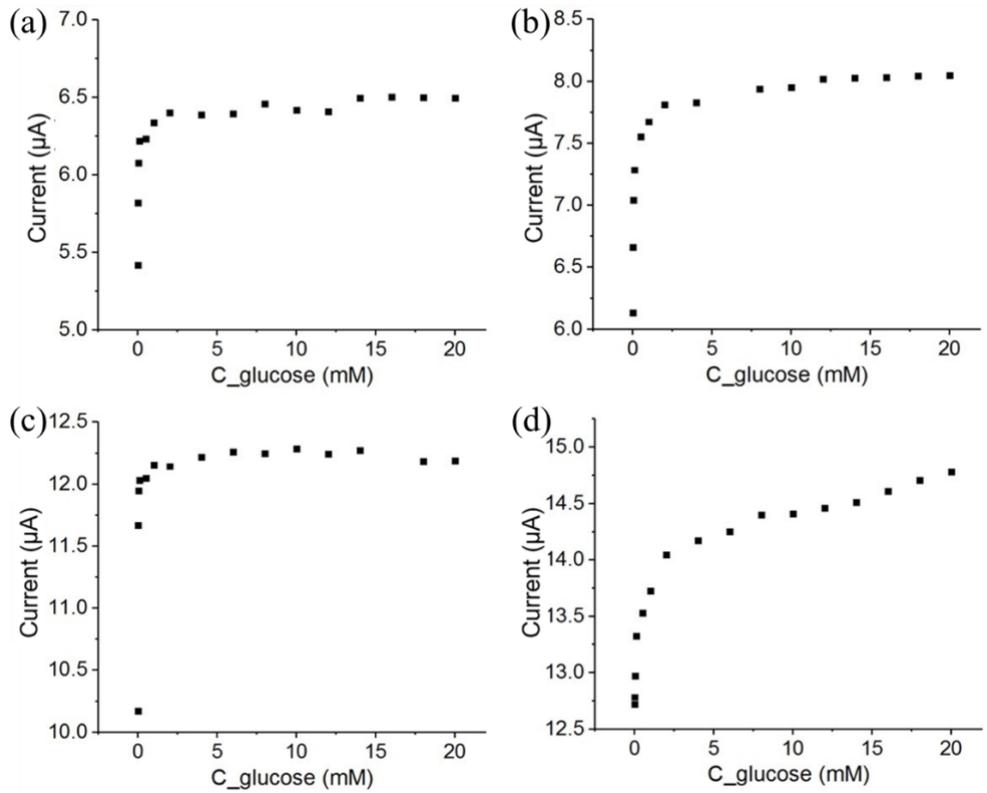

Fig. 4. Sensor response (drain current) of the 4 individual devices with the applied bias $V_{gs}$ = -0.5 V and at $V_{ds}$ = 0.1 V. The concentration of glucose was varied from 0.01 to 20 mM.

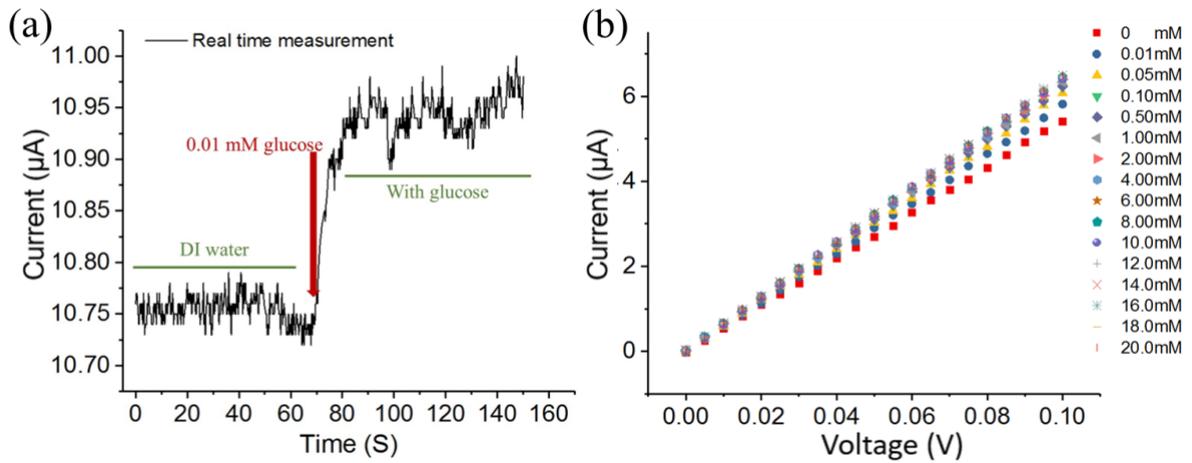

Fig. 5. Real time response of the CNTFET biosensor. (a) Biosensor response at 0.01 mM glucose. (b) Biosensor response with different concentrations of glucose (0.01-20 mM). $V_{gs}$ = -0.5 V and at $V_{ds}$ = 0.1 V.

Table 1. Current change (ΔI) with various glucose concentrations (C_glucose) for 3 sets of 18 different devices with the applied bias $V_{gs}$ = -1.5 V and at $V_{ds}$ = 1 V.

| Sample | C_glucose mM | ΔI = (I2-I1) |
|---|---|---|
| Set 1, 6 devices | 0.01 | 3.10E-07 |
| | 0.01 | 4.01E-07 |
| | 0.01 | 9.21E-07 |
| | 0.01 | 1.15E-07 |
| | 0.01 | 9.58E-07 |
| | 0.01 | 1.16E-06 |
| Set 2, 6 devices | 0.05 | 5.65E-06 |
| | 0.05 | 6.57E-06 |
| | 0.05 | 3.81E-06 |
| | 0.05 | 7.17E-08 |
| | 0.05 | 2.35E-07 |
| | 0.05 | 5.94E-07 |
| Set 3, 6 devices | 0.1 | 1.64E-06 |
| | 0.1 | 7.36E-06 |
| | 0.1 | 1.19E-06 |
| | 0.1 | 3.37E-07 |
| | 0.1 | 8.80E-07 |
| | 0.1 | 1.63E-06 |